\documentclass[twocolumn]{IEEEtran}
\usepackage[final]{graphicx}

\usepackage{amsmath,epsfig,amssymb,verbatim,amsopn,cite,multirow}
\usepackage{balance}
\usepackage{multirow}
\usepackage{footnote}
\usepackage{algorithm}
\usepackage{algpseudocode}
\usepackage[usenames,dvipsnames]{color}
\usepackage[all]{xy}  %%% used to make block diagram
\usepackage{url}
\usepackage{subcaption}
\usepackage{graphicx}
\usepackage{array}
\usepackage{amsthm}

\usepackage[nodisplayskipstretch]{setspace}
  {\proof}{\proofend}

\newcommand{\qa}{{\bf a}}

\newcommand{\qe}{{\bf e}}

\newcommand{\qg}{{\bf g}}
\newcommand{\qh}{{\bf h}}

\newcommand{\qs}{{\bf s}}

\newcommand{\qw}{{\bf w}}

\newcommand{\qE}{{\bf E}}

\newcommand{\qG}{{\bf G}}

\newcommand{\qI}{{\bf I}}

\newcommand{\sym}{ \mathrm{i} }
\newcommand{\mr}{ \mathrm{d} }
\newcommand{\zf}{ \mathrm{c} }
\newcommand{\wsc}{\qw^{\zf}}
\newcommand{\wsd}{\qw^{\mr}}
\newcommand{\kcset}{\mathcal{K}_{\zf}}  
\newcommand{\kdset}{\mathcal{K}_{\mr}} 
\newcommand{\Ex}{\mathbb{E}}

\newcommand{\Kmax}{K_{\mathrm{max}}}
\newcommand{\mset}{\mathcal{M}}

\newcommand{\Kset}{\mathcal {K}}

\newcommand{\tauu}{\tau_{\mathrm{u}}}

\newcommand{\Kc}{\mathcal{K}_c}
\newcommand{\Kd}{\mathcal{K}_d}
\newcommand{\SEc}{\mathcal{S}^c_k}
\newcommand{\SEd}{\mathcal{S}^d_k}

\newcommand{\etamtd}{\eta_{mt}^\mr}

\newcommand{\gamamkd}{\gamma_{mk}}
\newcommand{\gamamkc}{\gamma_{mk}}
\newcommand{\betamkd}{\beta_{mk}}
\newcommand{\betamkc}{\beta_{mk}}

\newcommand{\tkc}{t_k^{\mathrm{c}}}
\newcommand{\tkd}{t_k^{\mathrm{d}}}

\DeclareMathOperator{\C}{\mathbb{C}}

\newcommand{\etatc}{\eta_{t}^{\mathrm{c}}}
\newcommand{\etakc}{\eta_{k}^{\mathrm{c}}}

\title{\fontsize{0.83cm}{1cm}\selectfont  Hybrid Centralized-Distributed  Precoding  in Fronthaul-Constrained CF-mMIMO Systems}
\author{Zahra Mobini,~\IEEEmembership{Senior Member,~IEEE,} Hien Quoc Ngo,~\IEEEmembership{Fellow,~IEEE}, Ardavan Rahimian,~\IEEEmembership{Senior~Member,~IEEE}, Anvar Tukmanov,~\IEEEmembership{Senior~Member,~IEEE}, David Townend, Michail Matthaiou,~\IEEEmembership{Fellow,~IEEE}, and \\Simon L. Cotton,~\IEEEmembership{Fellow,~IEEE}}

\allowdisplaybreaks
\allowdisplaybreaks
\begin{document}

\bstctlcite{IEEEexample:BSTcontrol}
\maketitle

\begin{abstract} We investigate a fronthaul-limited cell-free massive multiple-input multiple-output (CF-mMIMO) system and propose a hybrid centralized-distributed  precoding strategy that dynamically adapts to varying fronthaul and spectral efficiency (SE) requirements. The proposed approach divides users into two groups: one served by centralized precoding and the other by distributed precoding. We formulate a novel optimization problem for user grouping and power control aimed at maximizing the sum SE, subject to fronthaul and per-access point (AP) power constraints. To tackle the problem,  we transform it into a tractable form and propose efficient solution algorithms. Numerical results confirm the hybrid scheme's versatility and superior performance, consistently outperforming fully centralized and distributed approaches across diverse system configurations.

\let\thefootnote\relax\footnotetext{
 This work was supported by the U.K. Engineering and Physical Sciences
Research Council (EPSRC) grant (EP/X04047X/2) for TITAN Telecoms
Hub.

%The work of Z.~Mobini and  H.~Q.~Ngo
 %was supported by the U.K. Research and Innovation Future
%Leaders Fellowships under Grant MR/X010635/1. 
Z.~Mobini is with the Department
of Electrical and Electronic Engineering, The University of Manchester,
Manchester M13 9PL, U.K., and also with the Centre for Wireless Innovation (CWI), Queen’s University Belfast, BT3 9DT Belfast, U.K. (e-mail: zahra.mobini@manchester.ac.uk).

H.~Q.~Ngo, M. Matthaiou, and S. L. Cotton are with the Centre for Wireless Innovation (CWI), Queen’s University Belfast, BT3 9DT Belfast, U.K. (email:\{hien.ngo,  m.matthaiou, simon.cotton\}@qub.ac.uk).

A. Rahimian is with the School of Engineering, Ulster University, Belfast, BT15 1AP, U.K. (e-mail: a.rahimian@ulster.ac.uk).

A. Tukmanov and D. Townend are with British Telecommunications PLC, IP53RE Ipswich, U.K. (e-mail:\{anvar.tukmanov, dave.townend\}@bt.com).
}
\end{abstract}
%==============================================================================
%========================================================\section{Introduction}
%==========================================================================
%============================================================
\vspace{-0.4em}
%============================================================
\vspace{-1em}
\section{Introduction}
Cell-free massive MIMO (CF-mMIMO) is a leading candidate for next-generation  networks, boosting spectral efficiency (SE), energy efficiency (EE), and connectivity~\cite{Mohammadali:TCOM:2024,Ngo:PROC:2024}. It employs multiple distributed access points (APs) coordinated by a central processing unit (CPU) to serve users simultaneously but faces challenges from limited fronthaul capacity due to overhead in data, channel state information (CSI), and precoding vector transmission. Addressing this requires solutions that balance performance with fronthaul efficiency~\cite{Mohammadali:survey:2024}.

In line with recent trends towards disaggregated and software-defined radio access networks (RANs), exemplified by initiatives like the open RAN (O-RAN) Alliance, CF-mMIMO systems can also leverage a functional split architecture~\cite{Demir:2024:JSAC}. In this setup, an AP consists of a remote radio head (RRH) for radio frequency (RF) processing and a baseband unit (BBU) for signal processing functions (such as channel estimation, precoding, encoding and modulation), which is further split into  baseband high unit (BBH) and  baseband low unit (BBL).  BBLs are connected to their associated BBHs via fronthaul links to transmit information, such as precoding vectors and data signals, while BBHs are connected through backhaul links for information exchange between the APs. The allocation of tasks/functions between the BBH and BBL introduces a trade-off between the system performance  and the fronthaul requirements. Allocating more functions to the BBH improves coordination but increases fronthaul overhead, while moving functions to the BBL reduces fronthaul load at the cost of coordination and SE. In O-RAN, functional split 7c supports distributed precoding at the BBL to lower fronthaul load, whereas split 7b centralizes precoding at the BBH~\cite{Romero:2024:ISWCS}.

Conventional CF-mMIMO systems use either functional split 7b or 7c. Studies \cite{hien:2017:wcom, Hao:IOT:2024} examined non-cooperative precoding methods, like matched filtering (MF) and local zero-forcing (ZF), with split 7c to reduce fronthaul  overhead and complexity. 
Moreover, recent distributed precoding methods include neural-inspired training without CSI sharing~\cite{Dai:2025:TCOM}, federated learning~\cite{Wang:2023:ojcom}, and decentralized cluster-based precoding using only local CSI within AP clusters to reduce backhaul overhead and complexity \cite{shi:2023:tvt}.\footnote{While some literature refers to the AP–CPU link as the “fronthaul,” in our work, we use the term “fronthaul” to denote the link between the BBL–BBH  in line with functional split architectures. This numerology is adopted without loss of generality. }
Conversely, centralized precoding in split 7b enables full cooperation among APs (or BBHs) for better performance \cite{nayebi:2017:wcom} but demands extensive CSI exchange, higher complexity, and greater fronthaul resources as user numbers increase. Designing systems that balance different functional approaches effectively is challenging, driving the need for innovative solutions.

This paper proposes a hybrid centralized-distributed precoding strategy that balances fronthaul capacity and SE by dividing users into two groups: one using global CSI with precoding at the BBHs, and the other using local CSI with precoding at the BBLs. This approach supports scalable, cost-effective, and flexible transitions toward centralized architectures in future wireless networks. The key contributions of this letter include:
%====
\begin{itemize}
\item We develop an analytical framework for fronthaul-limited CF-mMIMO systems with hybrid centralized-distributed  precoding, and formulate an optimization problem for user grouping and power control to maximize the sum SE under fronthaul and per-AP power constraints.

\item To solve the challenging optimization problem, we decompose the problem into two sub-problems: user grouping and power optimization, and develop a successive convex approximation (SCA)-based algorithm for power optimization and a K-means approach for user grouping. 

\item  Numerical results demonstrate that our optimization approach significantly outperforms heuristic methods. Moreover, they confirm that the hybrid scheme is well-suited for a wide range of CF-mMIMO applications.
\end{itemize}

%=============================
\textit{Notation:} %We use bold upper case letters to denote matrices, and lower case letters to denote vectors.  
The superscripts $(\cdot)^H$ and $\mathbb{E}\{\cdot\}$ stand for the conjugate-transpose and statistical expectation. A circular symmetric complex Gaussian distribution with variance  $\sigma^2$ is denoted by $\mathcal{CN}(0,\sigma^2)$, while $\mathbf{I}_N$ is the $N \times  N$ identity matrix.  %==============================================================================
\vspace{-2em}
\section{System Model}
%==============================================================================
We study a CF-mMIMO system consisting of $M$  APs and $K$ users. The APs are represented by the set $\mathcal{M} = \{1, \ldots, M\}$, and the users by the set $\mathcal{K} = \{1, \ldots, K\}$. Each AP is equipped with $L$ antennas, whereas the users are equipped with a single antenna each. A frequency-flat, slow fading channel  is assumed for every orthogonal frequency-division multiplexing (OFDM) subcarrier. For simplicity, we omit the subcarrier index. The  channel vector between the $m$-th AP and the $k$-th user is denoted by $\qg_{mk} \in \mathbb{C}^{L \times 1}$, and it is modeled as 
$ {\qg}_{mk} = \beta ^{1/2}_{mk}\qh_{mk}$,
 %==========================
where $\beta_{mk}$ represents the large-scale fading coefficient, which accounts for both path-loss and shadowing effects. Moreover, $ {\qh}_{mk} \in \mathbb{C}^{L \times 1}$ denotes the small-scale fading vector, with its entries being independent and identically distributed (i.i.d.) random variables following a $\mathcal{CN}(0, 1)$ distribution. We consider functional split 7c for the BBU, where the BBLs are equipped with the capability to perform precoding operations. Each BBH has full (global) CSI knowledge, that is BBH $m$ knows estimates of $\{\qg_{mk}\}$, $m \in \mset, k\in \Kset$. On the other hand, BBLs have local CSI knowledge, i.e., BBL $m$ knows the estimates of $\{\qg_{mk}\}$, for $k\in \Kset$.  
The system utilizes a time division duplex (TDD) protocol. Our focus is on downlink transmission, where each coherence interval of length $\tau$ is split into an uplink training phase and a downlink data transmission phase.   

In the uplink training phase, all users   transmit their pairwisely orthogonal pilot sequences of
length $\tau_{\mathrm{u}}$ to all the APs, which requires $\tau_{\mathrm{u}}\geq K$. These pilot signals enable each AP  to estimate the channels for all  users. The channel estimation process is carried out   at the AP’s  BBH, where the received pilot signals are processed to extract the  CSI. Using the minimum mean square error (MMSE) estimation method, the MMSE estimate of  ${\qg}_{mk}$  is  $\hat{\qg}_{mk} \sim \mathcal{CN}({\bf 0}, \gamma_{mk}\qI_L)$ where  
$		\gamma_{mk}\triangleq 
  {\tauu \rho_{u} \beta_{mk}^{2}}/{(\tau_{u} \rho_{u} \beta_{mk}+1}), $
while $\rho_{u}$ denotes the normalized  power of each pilot symbol.
%---------------------------------------------
%--------------------------------
%================
\vspace{-1em}
\subsection{Proposed Hybrid Centralized-Distributed   Precoding} \label{sec:Downlink Data Transmission}
Conventionally, precoding is either fully centralized at BBHs or fully distributed at BBLs. Our proposed hybrid scheme combines both approaches by processing some precoding vectors at BBHs and others at BBLs, achieving a balance between performance and fronthaul requirements.
Specifically, $K$ users in the system are divided into two precoding groups: $\kcset$ and $\kdset$ with cardinalities $K_{\zf}$ and $K_{\mr}$, respectively, where  $\kcset \cap \kdset=\emptyset$ and $\kcset  \cup \kdset=\Kset$.   The user $k$ in group $\kcset$ is served using global CSI, where the associated  precoding vectors $\wsc_{mk}$, $k\in \kcset, m \in \mset$, are computed at the BBHs.
 The user $k$ in group $\kdset$ is served  using local CSI, where the associated precoding vectors $\wsd_{mk}$, $k\in \kdset, m \in \mset$, are computed at the BBLs. This eliminates the need for precoding vector exchange between the BBL and its corresponding BBH.
 %=========
%-------------------------------------------------------------------------
Therefore, the $L \times 1 $ precoded signal to be transmitted from the $m$-AP to user $k$    can be expressed as
%==========================
\begin{equation}\label{eq:Sm}
	\qs_{m}=\!\sum _{k\in \kcset}\!\!\sqrt{\eta_{mk}^\zf} \wsc_{mk} x_{k}+\sum _{k\in \kdset}\! \! \sqrt{\eta_{mk}^\mr}\!   \wsd_{mk} x_{k},  \forall m\in \mset,
\end{equation}
%==========================
where   $\eta_{mk}^\zf (\eta_{mk}^\mr)$  represents  the $k$-th user power control coefficient, $k \in \kcset$ ($k \in \kdset$), and $x_{k}$ represents the information symbol  for the $k$-th user  with
 $\mathbb{E}\{|x_{k}|^2 \} = 1$.  The  $k$-th user in $\kcset$ ($\kdset$) receives signal $r_{k}^{\zf}$  ($r_{k}^{\mr}$) as
%%%%%%%%%%%%%%%%%%%%%%%%%%%%%%%%%5
%===================
\begin{align}\label{eq:received_r}
&r_{k}^\sym
=
\sum _{m\in \mset}\!\!
 \sqrt{\eta_{mk}^\sym}
    \qg^{H}_{mk} \qw_{mk}^\sym x_{k}+\!\!\sum _{m\in \mset} \sum _{\substack {k'\in\kcset,   k' \neq k }}\!\!\!\!\!\!\!\!\!\!\sqrt{\eta_{mk'}^\zf}\qg^{H}_{mk}\nonumber\\
&
 \times\wsc_{mk'}x_{k'}\!+\!\!\!\sum _{m\in \mset}\sum _{\substack {k'\in\kdset,  k' \neq k }}\!\!\!\!\!\!
\sqrt{\eta_{mk'}^\mr}
\qg^{H}_{mk} \wsd_{mk'}x_{k'} + n_k,
\end{align}
%==========================
where $ \sym \in \{\zf, \mr\}$ and $n_k\sim \mathcal{CN} (0, \sigma^2)$ is the  additive noise.   The first term in~\eqref{eq:received_r} describes the desired signal part, while the second and third terms account for the inter-user interference.
%============================================================================================

%============================================================================================
We use centralized ZF design to obtain the precoding vector $\wsc_{mk}$  for $k \in \kcset$ and  local ZF design  to obtain $\wsd_{mk}$  for $k \in \kdset$~\footnote{We adopt  ZF due to its near-optimality and efficiency in CF-mMIMO~\cite{Ngo:PROC:2024}, while fronthaul constraints are incorporated within our optimization framework in the next section.}. Let $\widehat {\qG}^\zf$ be an $ML\times K_\zf$ collective channel estimation matrix for all the APs and the corresponding users  in $\kcset$. More specifically, $\widehat {\qG}^\zf=[(\widehat {\qG}_{1}^\zf)^H,\cdots,(\widehat {\qG}_{M}^\zf)^H]^H$ consists of  $M$ channel estimates of dimension $L\times K_\zf$ each, where $\widehat {\qG}_{m}^\zf$ is the estimate of the channel matrix from AP $m$ and the users in set $\kcset$, i.e., $\widehat {\qG}_m^\zf=[\hat {\qg}_{mk}: k \in \kcset]$.
%========================
Let user $k$ correspond to the $i$-th element of set $\kcset$, $i \in \{1,\cdots,K_\zf\}$ and   $\qa_m=[(m-1)L+1,\cdots,mL]$. Then, the channel estimate vector between the
AP $m$ and user $k \in \kcset$ can be expressed in terms of $\widehat {\qG}^\zf$ as
$\hat {\qg}_{mk}=\qE_m\widehat {\qG}^\zf\qe_i$,
where $\qE_m=[\qe_{\qa_m(1)},\cdots,\qe_{\qa_m(L)}]^H \in \C^{L\times ML}$ with $\qe_{\qa_m(\ell)}$ shows  the $\qa_m(\ell)$-th column of $\qI_{ML}$, while $\qe_i$ is the $i$-th column of $\qI_{K_\zf}$. Accordingly,  the precoding vector constructed for AP $m$ to transmit to  user $k \in \kcset$ can be written as 
$\qw_{mk}^\zf= \qE_m
\widehat {\qG}^\zf \big(({\widehat {\qG}^{\zf})^H \widehat {\qG}^\zf }\big)^{-1}\qe_i, \forall k \in \kcset.$
 %============
Now, let user $k$ correspond to the $j$-th element of set $\kdset$, $j \in \{1,\ldots,K_\mr\}$. Then, we define $\boldsymbol{\pi}_j$ as the $j$-th column of $\qI_{K_\mr}$. The precoding vector constructed for AP $m$ to transmit to  user $k \in \kdset$ can be calculated as
%==============================
\begin{equation}\label{eq:Q}
\qw_{mk}^\mr=\frac{\widehat{\qG}_{m}^\mr((\widehat{\qG}_{m}^\mr)^H\widehat{\qG}_{m}^\mr)^{-1}\boldsymbol{\pi}_j}{\sqrt{\mathbb{E}\{\|\widehat{\qG}_{m}^\mr((\widehat{\qG}_{m}^\mr)^H\widehat{\qG}_{m}^\mr)^{-1}\boldsymbol{\pi}_j\|^2\}}}, \forall k \in \kdset,
\end{equation}
%============
where $\widehat {\qG}_m^\mr=[\hat {\qg}_{mk}: k \in \kdset]$.
 %===========
It is worth noting that, for the users in group $\kcset$, the second term in~\eqref{eq:received_r}, which represents intra-group interference, can be significantly reduced. However, for the users in $ \kdset$, an AP can only suppress its own interference and cannot mitigate interference from other APs. In addition,  as discussed in~\cite{nayebi:2017:wcom}, for the users in set $\kcset$,  the power allocation coefficients should only be a function of $k$, which means that $\eta_{mk}^\zf=\eta_{m'k}^\zf=\eta_{k}^\zf$, $\forall k \in \kcset$.

%=====================================================================
\vspace{-1.5em}
%----------------------------------------------------------------------
\subsection{Downlink Spectral Efficiency and Fronthaul Requirements}
To detect $x_k$ from the received signal in~\eqref{eq:received_r}, the user is assumed to rely on statistical CSI. 
By using the use-and-then-forget bounding technique~\cite{Hien:JCOM:2021}, an achievable   SE for user $k$ in $\kcset$ ($\kdset$), %$\mathrm{SE}_k^{\zf}$ ($\mathrm{SE}_k^{\mr}$),  
is 
 $\mathrm{SE}_k^{\sym}=\big(1-\frac{\tau_\mathrm{u}}{\tau}\big)\log_2(1+\mathrm{SINR}^{\sym}_k),$
 where    $\mathrm{SINR}^{\sym}_k$ can be calculated as 
 %=================
\begin{align}
~\label{eq:SINRk_2}
 &\mathrm{SINR}_k^{\zf}=\nonumber\\
 &\frac{\eta_k^\zf}{\sum\limits_{t\in \kcset}\eta_{t}^\zf\!\!\sum\limits_{m \in \mset}\mu_{mt}(\beta_{mk}-\gamma_{mk})\! +\!\!\! \sum\limits_{t \in \kdset}\sum\limits_{m \in \mset}\eta_{mt}^\mr\beta_{mk}\!+\!1},
 \end{align}
%======
\begin{align}
~\label{eq:SINRk_3}
 &\mathrm{SINR}_k^{\mr}=\nonumber\\
 &\frac{(L-K_\mr)\Big(\sum_{m \in \mset}\sqrt{\eta_{mk}^\mr\gamma_{mk}}\Big)^2}{\sum\limits_{t\in \kcset}\eta_{t}^\zf\!\!\sum\limits_{m \in \mset}\mu_{mt}\beta_{mk}+\!\!\! \sum\limits_{t \in \kdset}\sum\limits_{m \in \mset}\eta_{mt}^\mr(\beta_{mk}-\gamma_{mk})\!+\!1},
\end{align}
%======
 with $\mu_{mt}=\Ex\big\{ \|\qw_{mt}^\zf \|^2\big\}$.
 
Now, we formulate the total fronthaul requirement at each AP for transmission from BBH to BBL which consists of two parts: 1) 
fronthaul requirement  for downlink data transmission, $\mathrm{FH}_{m,\mathrm{data}}$, and 2) fronthaul requirement for sending the precoding vectors, $\mathrm{FH}_{m,\mathrm{pr}}$. In a fronthaul limited CF-mMIMO system the  per-AP fronthaul  constraint can be written as
%===================
\begin{align}\label{eq:Kmax_cons}
&\mathrm{FH}_{m,\mathrm{data}}+\mathrm{FH}_{m,\mathrm{pr}} \leq \mathrm{FH}_{\mathrm{max}},~~~ \forall m \in \mset,
\end{align}
 %===============
where $\mathrm{FH}_{\mathrm{max}}$ is the per-AP maximum fronthaul capacity. More specifically, for each AP $m$, the fronthaul consumption for transmitting information symbols to all the users in $\Kset$, utilizing packed-based evolved common public radio interface (eCPRI) for the fronthaul  transmission, is given by $\mathrm{FH}_{m,\mathrm{data}}=(K_\zf+K_\mr)\alpha_1,$
 %================================
with 
$\alpha_1 \triangleq \frac{\log_2 (M_{\mathrm{order}}) N_{\mathrm{subcarrier}} N_{\mathrm{OFDM}}}{\mathrm{eCPR_{eff}}\mathrm{delay_{data}}},$
%================================
where $M_{\mathrm{order}}$ is the modulation cardinality, $N_{\mathrm{subcarrier}}$ is the number of OFDM subcarriers,  $\mathrm{delay}_{\mathrm{data}}$  shows the  transmit
delay for the data,   and $\mathrm{eCPR}_{\mathrm{eff}}$ is the efficiency of CPRI. In addition, the required fronthaul requirement for sending the precoding vectors to users in $\kcset$ can be written as 
$\mathrm{FH}_{m,\mathrm{pr}}=  K_\zf \alpha_2,$
%================================
with 
$\alpha_2 \triangleq \frac{2L  N_{\mathrm{bits}}N_{\mathrm{Gran}}}{\mathrm{eCPR_{eff}}\mathrm{delay_{pr}}} $,
%================================
where $N_{\mathrm{bits}}$ denotes the number of quantization bits, while $N_{\mathrm{Gran}}$ shows the precoding granularity, and $\mathrm{delay_{pr}}$  shows the transmit delay of the precoding weights.   Accordingly, in a fronthaul limited CF-mMIMO system with the  fronthaul constraint~\eqref{eq:Kmax_cons}, we must constrain the precoding traffic  to be fronthauled  by restricting the number of users  as
%===========================
\begin{align}\label{eq:Kmax}
K_\zf \alpha_2+ (K_\zf+K_\mr)\alpha_1 \leq  \mathrm{FH}_{\mathrm{max}}.
\end{align}
%================================
%=================================================================
%============================================================
%=================================================================
%============================================================

%============================================================
\vspace{-1em}
\section{User Grouping and Power Allocation}
In this section, we aim to optimize  the transmit power allocation coefficients and user grouping  for the maximization of the sum SE across all users, subject to per-user SE, per-AP fronthaul capacity, and maximum transmit power constraints. With~\eqref{eq:Sm}, the expected transmit signal power from AP $m$ is given by 
$\mathrm {\mathbb {E}}\{\qs_{m}^H \qs_{m}\}=\sum\nolimits_{k \in \kcset} \eta_k^\zf\mu_{mk}+\sum\nolimits_{k \in \kdset} \eta_{mk}^\mr$.
 %========================================
Accordingly,   the   optimization problem  can be formulated as
%---------------------
\begin{subequations}\label{eq:problem1}
\begin{alignat}{2}
&  \max _{\boldsymbol{\eta},\kcset,\kdset}          
&\quad& \sum\nolimits_{k\in\Kc}\mathrm{SE}_k^{\zf} +   \sum\nolimits_{k\in\Kd}\mathrm{SE}_k^{\mr}\label{eq:problem1:O1}\\
%----
&\hspace{2.5em}\text{s.t.} 
%----    
&         &       K_\zf \alpha_2+ (K_\zf+K_\mr)\alpha_1 \leq  \mathrm{FH}_{\mathrm{max}}, ~\label{eq:problem1:C1}\\
%----
&         &      &\mathrm{SE}_k^{\zf} \geq \SEc, \quad \mathrm{SE}_k^{\mr} \geq \SEd,~\label{eq:problem1:Czf}\\
%---
%----
&         &      &\sum\nolimits_{k \in \kcset} \eta_k^\zf\mu_{mk}+\sum\nolimits_{k \in \kdset} \eta_{mk}^\mr \leq \rho,~\label{eq:problem1:C2}
%----
\end{alignat}
\end{subequations}
%------------------------
where     $\SEc$ ($\SEd$) is the minimum SE required by the $k$-th user in $\kcset$ ($\kdset$) to guarantee the required quality of service (QoS) in the network, and $\rho$ is the maximum normalized transmit power at each AP. Moreover, $\boldsymbol {\eta }$ denotes  the set of all power control coefficients $\boldsymbol {\eta } \triangleq \{\eta_{k}^\zf: k \in \kcset\} \bigcup \{\eta_{mk}^\mr: (m,k) \in \mset \times \kdset\}$.  The problem in~\eqref{eq:problem1} is non-convex. Another challenge arises from the fact that the precoding vectors $\qw_{mk}^\zf$ are a non-convex function of $\kcset$. The problem~\eqref{eq:problem1} is NP-hard, making global optimization infeasible; thus, we separate the optimization variables and solve sub-problems using efficient algorithms.
%===============

\vspace{-0.6em}
%======
\subsection{User Grouping}\label{user grouping}
To divide users into two groups, $\kcset$ and $\kdset$, we adopt a modified K-means approach inspired by unsupervised learning. This scheme is well-suited to irregular user distributions or heterogeneous channel conditions, and adaptive without predefined thresholds. While not optimal, it offers a practical and tractable solution for user grouping within our joint optimization framework.
 In the context of CF-mMIMO, we use large-scale fading vector $\boldsymbol{\beta}_k=[\beta_{1k},\ldots,\beta_{Mk}]^T$  as the feature vector for user $k$~\cite{Riera:Globecom:2018}.
 The distance between two fingerprint vectors, $\boldsymbol{\beta}_k$
and $\boldsymbol{\beta}_{k'}$, is measured using the cosine distance, defined as $
    f_d(\boldsymbol{\beta}_k,\boldsymbol{\beta}_{k'})=1-\frac{\boldsymbol{\beta}_k^T\boldsymbol{\beta}_{k'}}{\|\boldsymbol{\beta}_k\|\|\boldsymbol{\beta}_{k'}\|}.
$ 
%=============
This metric
%ranges from $1$ (indicating orthogonality or no correlation) to $0$ (indicating identical vectors), with intermediate values
represents the degrees of similarity~\cite{Riera:Globecom:2018}. The reason for this grouping strategy is that users with cosine distance values close to  $0$ are likely to be geographically close and thus experience similar statistical channel conditions. Consequently, transmissions from the APs to these neighbouring users will contribute significantly to inter-user interference. To mitigate this, users exhibiting the highest correlation (cosine distance near $0$) are included in $\kcset$  and served using centralized  ZF, as this approach is more effective in managing interference. The remaining users, forming $\kdset$, are served using local ZF. This method effectively minimizes inter-user interference and enhances the   SE performance.

\vspace{-1em}
%-----------------------------
\subsection{Power Allocation Optimization Given  User Grouping}~\label{Power Allocation}
The power allocation problem with fixed user grouping $\kcset$ and $\kdset$ is expressed as
%======
%---------------------
\begin{subequations}\label{eq:opt:SE}
\begin{alignat}{2}
&\mathcal{P}.1: \max_{\boldsymbol{\eta}}        
&\quad& \sum\nolimits_{k\in\Kc}\mathrm{SE}_k^{\zf} + \sum\nolimits_{k\in\Kd} \mathrm{SE}_k^{\mr}~\label{eq:pq:obj}\\
%----
&\hspace{0.5em}\text{s.t.} 
%----    
&         &    \eqref{eq:problem1:Czf},%\eqref{eq:problem1:Cmr},
\eqref{eq:problem1:C2}. 
%----
\end{alignat}
\end{subequations}
%------------------------
The power optimization problem $\mathcal{P}.1$ in \eqref{eq:opt:SE} is inherently non-convex, making it challenging to solve directly. To address this complexity, we employ the SCA technique.\footnote{The application of the SCA-based approach here is non-trivial due to the need for convex surrogate functions that ensure convergence under practical system constraints within our hybrid precoding framework.} This approach 
tackles iteratively a sequence of convex approximations of the original problem.  By introducing slack variables 
$t_k^\zf$ and $t_k^\mr$, the problem  $\mathcal{P}.1$ can be reformulated into an equivalent form as
%---------------------
\vspace{0.5em}
\begin{subequations}\label{eq:opt:SE2}
\begin{alignat}{2}
&\mathcal{P}.2: \max_{\boldsymbol{\eta}}        
&\quad& \prod\nolimits_{k\in\Kc}(1+\tkc) + \prod\nolimits_{k\in\Kd} (1+\tkd)~\label{eq:SE2:obj}\\
%----
&\hspace{1.5em}\text{s.t.} 
%----    
&         &       \mathrm{SINR}_k^{\zf} \geq \tkc, ~\label{eq:SE2:ct1}\\
%----
&         &      &\mathrm{SINR}_k^{\mr} \geq \tkd,~\label{eq:SE2:ct2}\\
%---
&         &       &\log_2(1+\tkc)\geq \SEc,~\log_2(1+\tkd)\geq \SEd, ~\label{eq:SE2:ct3}\\
%----
%&         &      &\log_2(1+\tkd)\geq \SEd,~\label{eq:SE2:ct4}\\
%----
&         &      &\eqref{eq:problem1:C2}.
%----
\end{alignat}
\end{subequations}
%------------------------
%------------------------
The primary difficulty in solving $\mathcal{P}.2$ stems from the non-convexity of the constraints in \eqref{eq:SE2:ct1} and \eqref{eq:SE2:ct2}. As previously discussed, the SCA method is employed to address these non-convex constraints. To begin, we focus on the constraint in \eqref{eq:SE2:ct1}, which can be reformulated as
\vspace{-1em}
%=====
\begin{align}\label{eq:C1}
     \etakc\geq \!\!\sum_{t\in \kcset}\!\!u_{tk}\tkc\etatc  \! \!+ \! \sum_{t \in \kdset}\!\sum_{m \in \mset}\!\!\betamkc\tkc\etamtd+\tkc,~\forall k\in\mathcal{K}_c,
\end{align}
where $u_{tk}\triangleq \sum_{m \in \mset}\mu_{mt}(\betamkc-\gamamkc)$. Using the equality $4xy=(x+y)^2-(x-y)^2$, the  inequality~\eqref{eq:C1} can be re-written as
%====
\begin{align}
      4\etakc&\geq \sum_{t\in \kcset}u_{tk}\Big[(\tkc+\etatc)^2-(\tkc-\etatc)^2\Big]+ \sum_{t \in \kdset}\sum_{m \in \mset}\betamkc \nonumber\\
      &\times\Big[(\tkc+\etamtd)^2-(\tkc-\etamtd)^2\Big]+4\tkc.
\end{align}
%====
Now, let us introduce $(n)$ to denote
the value of the variables produced after $(n-1)$ iterations ($n\geq0$) in the SCA-based approach. Leveraging the principles of SCA and using the inequality $x^{2}\geq  x_{0}(2x-x_{0})$, the original constraint~\eqref{eq:SE2:ct1} can be approximated by the following convex formulation:
%======
\begin{align}\label{eq:SE2:ct1_f}
    4\etakc &\geq \sum_{t\in \kcset}u_{tk}\Big[(\tkc+\!\etatc)^2-v_{tk}^{(n)} (2v_{tk}\!-\!v_{tk}^{(n)})\Big]\! +\! \sum_{t \in \kdset}\sum_{m \in \mset}\nonumber\\
     &\betamkc\Big[(\tkc+\etamtd)^2-z_{m,tk}^{(n)} (2z_{m,tk}-z_{m,tk}^{(n)})\Big]+4\tkc,
\end{align}
%====
where $v_{tk} = \tkc-\etatc$ and $z_{m,tk}=\tkc-\etamtd$.

Let us now consider \eqref{eq:SE2:ct2}. The objective is to find a concave lower bound for the left-hand side of the inequality. To accomplish this, we note that the function 
$x^{2}/y$ is convex when  $y>0$, which allows us to derive the following inequality:
%-------
\vspace{-0.2em}
\begin{align}\label{eq:x2/y}
\frac{x^{2}}{y}\geq
% \frac{x_{0}^{2}}{y_{0}}+2\frac{x_{0}}{y_{0}}(x-x_{0})-\frac{x_{0}^{2}}{y_{0}^{2}}(y-y_{0})=
\frac{x_{0}}{y_{0}}\Bigl(2x-\frac{x_{0}}{y_{0}}y\Bigr),
\end{align}
%----- 
where the inequality is obtained by linearizing the function 
$x^{2}/y$ around the points  $x_{0}$ and $y_{0}$.  Accordingly, we have
%======
\begin{align}\label{{eq:x2/y}}
& \frac{\big(\sum_{m \in \mset}\sqrt{(L-K_\mr)\eta_{mk}^\mr\gamma_{mk}}\big)^2}{\tkd}  \geq \nonumber \\
 &~~q_k^{(n)}\Big(2\sum_{m \in \mset}\sqrt{(L-K_\mr)\eta_{mk}^\mr\gamamkd}-q_k^{(n)} \tkd\Big), 
\end{align}
%------
where $q_k^{(n)}=\frac{\sum_{m \in \mset}\sqrt{(L-K_\mr)(\eta_{mk}^\mr)^{(n)}\gamma_{mk}}}{(\tkd)^{(n)} }$. Therefore, by applying~\eqref{eq:x2/y},  \eqref{eq:SE2:ct2} can be approximated as
%======
\begin{align}\label{eq:SE2:ct2_f}
&q_k^{(n)}\Big(2\sum_{m \in \mset}\sqrt{(L-K_\mr)\eta_{mk}^\mr\gamamkd}-q_k^{(n)} \tkd\Big)\geq \nonumber\\
&\sum_{t\in \kcset}\!\!\etatc\sum_{m \in \mset}\!\mu_{mt}\betamkd\!+\! \sum_{t \in \kdset}\!\sum_{m \in \mset}\etamtd(\betamkd-\gamamkd)\!+\!1.
\end{align}
%======
It is worth mentioning that constraint~\eqref{eq:SE2:ct2_f} is convex, which can be attributed to the concavity of 
$\sqrt{\eta_{mk}^\mr}$. Building upon the preceding analysis, we can derive the following approximate convex problem:
%---------------------
\begin{subequations}\label{eq:opt:SE3}
\begin{alignat}{2}
&\mathcal{P}.3: \max_{\boldsymbol{\eta}}        
&\quad& \prod\nolimits_{k\in\Kc}(1+\tkc) + \prod\nolimits_{k\in\Kd} (1+\tkd)~\label{eq:SE2:obj}\\
%----
&\hspace{0.5em}\text{s.t.} 
%----    
&         &  \eqref{eq:problem1:C2},  \eqref{eq:SE2:ct3},
%\eqref{eq:SE2:ct4},
\eqref{eq:SE2:ct1_f}, \eqref{eq:SE2:ct2_f}. ~\label{eq:SE3:ct31}
%---
%----
%----
\end{alignat}
\end{subequations}    
%==============
Existing convex optimization tools, such as CVX, %~\cite{cvx} 
can efficiently solve the convex optimization problem~\eqref{eq:opt:SE3}.
Our user grouping and power allocation rely on large-scale fading, leveraging channel hardening in CF-mMIMO systems~\cite{hien:2017:wcom } to enable efficient, scalable optimization based on statistical CSI across subcarriers and time frames~\cite{ZAHRA:msp:2025}.
%===========================================
%----------------------------------------------------

\begin{table}[t] 
\small
\centering
\caption{ Precoding schemes: Computational complexity and fronthaul usage}
\begin{tabular}{|c|c|c|}
\hline
\textbf{} & Computational Complexity & $\mathrm{FH}_{m,\mathrm{pr}}$ (Gbps) \\
\hline
\raisebox{1.2ex}{\textbf{Hybrid}} & 
\shortstack{$\mathcal{O}(MLK_c^2 + K_c^3)$\\$+ \mathcal{O}(LK_d^2 + K_d^3)$} & 
\raisebox{1.2ex}{$K_\zf \alpha_2$} \\
\hline
\textbf{Centralized} & $\mathcal{O}(MLK^2 + K^3)$ & $K \alpha_2$ \\
\hline
\textbf{Distributed} & $\mathcal{O}(LK^2 + K^3)$ & 0 \\
\hline
\end{tabular}
\label{tab:complex}
\end{table}

\vspace{-0.8em}
\section{Numerical Results}
We analyze a CF-mMIMO network with  $M$ APs and  $K$ users randomly distributed in a $2 \times 2$ km${}^2$ area using a wrapped-around topology, with $\tau = 2000$ samples and $\SEc=\SEd=1$ bit/s/Hz.
Each AP has a maximum transmit power of 
$500$ mW for training pilot sequences and $1$ W for data transmission, $\sigma^2_n=-92$~dBm and $K=20$. In addition, the fronthaul parameters are chosen as  $N_{\mathrm{subcarrier}}=3264$ subcarriers, $N_{\mathrm{Gran}} =136$, $\mathrm{eCPR_{eff}}=0.85, \mathrm{delay_{\mathrm{pr}}} (\mathrm{delay_{data}}) =     2\times10^{-4}(5\times10^{-4}) s $, $N_{\mathrm{OFDM}}  =14$,   $N_{\mathrm{bits}}=16$, and $M_\mathrm{{order}}=64$.  The large-scale fading and path loss are modeled according to the framework presented in~\cite{hien:2017:wcom}.  We consider two baseline schemes for comparison:   1) \textbf{Centralized}:  Centralized ZF is used for all  users.   2) \textbf{Distributed}: Local ZF is used for all  users.   The maximum number of users that can be served by the centralized (distributed) scheme, $\Kmax^\zf$ ($\Kmax^\mr$), can be determined by setting $K_\mr=0$ ($K_\zf=0$) in~\eqref{eq:Kmax}, while $\Kmax^\mr >\Kmax^\zf$. The user grouping algorithm is implemented as follows: For a given value of $\mathrm{FH}_{\mathrm{max}}$, we first calculate $\Kmax^\zf$. Then, we vary $K_\zf$ from 0 to $\Kmax^\zf$. For each value of $K_\zf$, the K-means approach is applied to perform user grouping, and the sum SE is computed for the resulting grouping, while $\Kmax^\mr$ is determined according to~\eqref{eq:Kmax}. Finally, we select the user grouping   that yields the maximum sum SE.    
The computational complexity of the precoding design and the fronthaul usage $\mathrm{FH}_{m,\mathrm{pr}}$ for different schemes are compared in Table~\ref{tab:complex}.

%%%%%%%%%%%%%%%%%%%%%%%%%%%%%%%%% format by Mohammadi 2
\begin{figure*}[t]
\vspace{-0.5cm}
    \centering
    % First figure
    \begin{minipage}[t]{0.31\textwidth}
        \centering
          \includegraphics[trim=0 0cm 0cm 0cm,clip,width=1.12\textwidth]{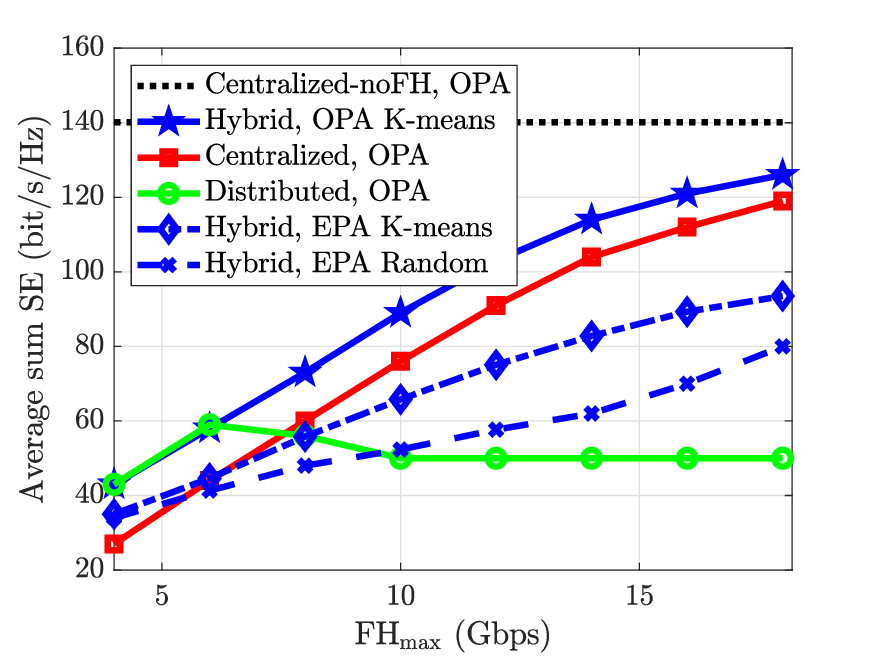}
          \vspace{-1.7em}
        \caption{\small Average sum SE  versus  $\mathrm{FH_{max}}$   ($M=20$,   $L=14$).\normalsize}
    \label{fig1}
    \end{minipage}
    \hfill
    % Second figure
    \begin{minipage}[t]{0.31\textwidth}
        \centering
          \includegraphics[trim=0 0cm 0cm 0cm,clip,width=1.12\textwidth]{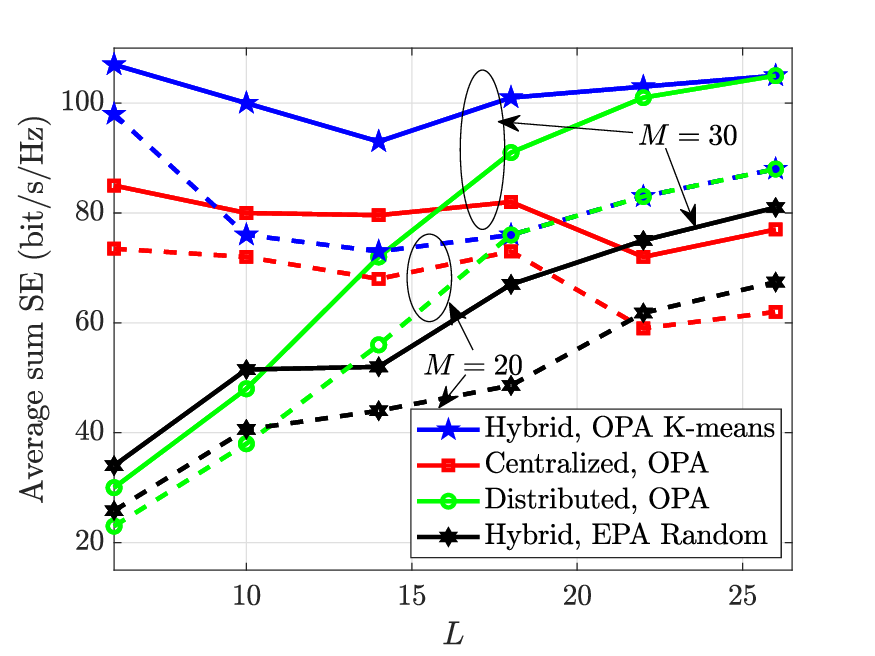}
          \vspace{-1.7em}
        \caption{\small  Average sum SE  versus  $L$   ($\mathrm{FH_{max}}=8$ Gbps). \normalsize}
        \label{fig2}
    \end{minipage}
    \hfill
    % Third figure
    \begin{minipage}[t]{0.31\textwidth}
        \centering
            \includegraphics[trim=0 0cm 0cm 0cm,clip,width=1.12\textwidth]{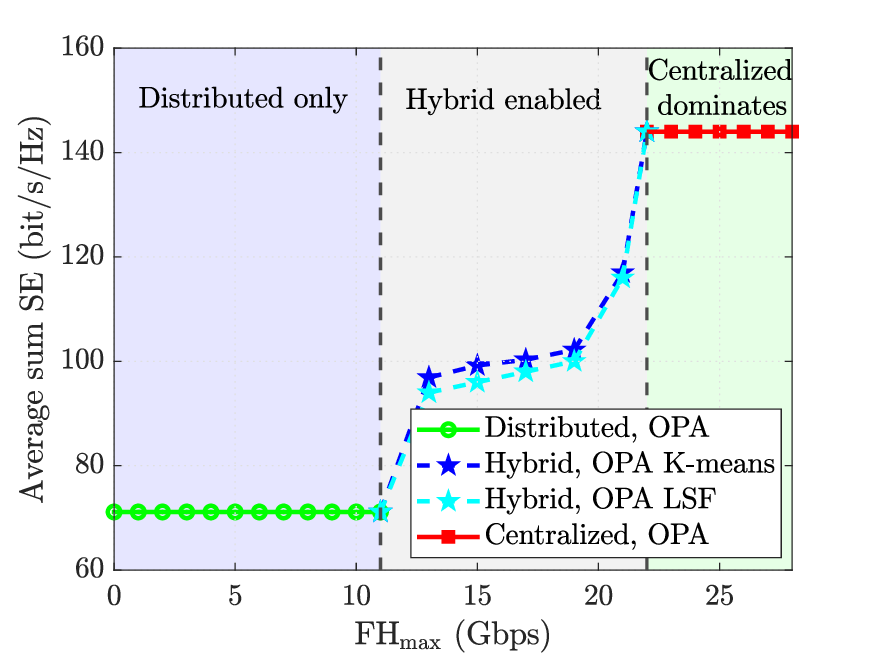}
            \vspace{-1.7em}
        \caption{\small  Average sum SE  versus $\mathrm{FH_{max}}$   ($M\!=\!20,L\!=\!21$,\! $M_\mathrm{{order}}=32$).\normalsize}
        \label{fig3}
    \end{minipage}
\vspace{-1.8em}
\end{figure*}
%%%%%%%%%%%%%%%%%%%%%%%%%%%%%%%%%

In Fig.~\ref{fig1},  we evaluate the performance of
the proposed hybrid centralized-distributed  precoding approach   in the fronthaul-aware   CF-mMIMO system. 
%===
%===========================================
%----------------------------------------------------
\textbf{OPA} and \textbf{EPA} refer to the proposed and equal power allocation schemes, respectively, while \textbf{noFH} denotes the case without fronthaul capacity limitations.
Numerical results lead to the following conclusions. 1) Under fully centralized precoding design, increasing 
$\mathrm{FH_{max}}$ naturally leads to better performance, as it allows a higher number of users, $\Kmax^\zf$, to be served,  and to enhanced interference cancellation due to the centralized  design; 2) On the other hand, in fully distributed precoding, there is a trade-off between per-AP precoding capability and increasing $\mathrm{FH_{max}}$ ($\Kmax^\mr$). Increasing $\Kmax^\mr$ without cooperative interference cancellation harms performance. Here, $\mathrm{FH_{max}}=6$ Gbps suffices, and raising it to 10 Gbps does not improve sum-SE because distributed beamforming only mitigates interference to its own users, not from other APs, which grows with $\mathrm{FH_{max}}$ ($\Kmax^\mr$). Also, $\Kmax^\mr$ must satisfy $\Kmax^\mr \leq L-1$; with $L=14$ antennas, the maximum $\Kmax^\mr=13$, corresponding to $\mathrm{FH_{max}}=9$ Gbps. Beyond this, increasing $\mathrm{FH_{max}}$ does not affect distributed precoding performance; 3) The proposed power allocation and user grouping
solution, yields significant sum-SE  performance
gain against \textbf{EPA} and random user grouping. More specifically,   the K-means-based user grouping provides
a performance gain of $34\%$, while the K-means based user grouping together with the power allocation can
provide a performance gain of  $83\%$ when $\mathrm{FH_{max}}=12$ Gbps. This highlights
the advantage of our proposed solution; 4) Our proposed hybrid precoding scheme offers significant advantages over both centralized and distributed approaches. By splitting users into two groups—one managed with distributed processing to reduce fronthaul demands and the other with centralized processing for better interference cancellation—it combines the strengths of both methods. This allows the hybrid scheme to achieve superior performance across all $\mathrm{FH_{max}}$ values. Its flexibility balances computational efficiency and communication overhead, making it well-suited for diverse CF-mMIMO applications with varying fronthaul needs.

% Figure~\ref{fig2} illustrates the performance of
% different precoding schemes as a function of the number of antennas per AP, $L$, for different number of APs, $M$. The performance of the distributed precoding scheme   increases sharply with $L$ due to the higher number of antennas. However, this trend is not observed for the centralized and hybrid precoding schemes. The key reason for this is that increasing   $L$ has two  contradicting effects on the system performance:
% 1) Positive impact: it enhances the macro-diversity gain;  2) Negative impact: it increases the
%  traffic that must be fronthauled betwee n BBHs and BBLs to transmit the precoding vectors, $\mathrm{FH}_{m,\mathrm{pr}}$, thus reducing $K_\zf+K_\mr$  ($\Kmax^\zf$)  in the CF-mMIMO system relying on hybrid (centralized) precoding scheme.
% As observed, when $M=20$, in the regime of $L \leq 18$, centralized precoding outperforms the distributed scheme. However, beyond this point, the performance of the distributed scheme surpasses that of the centralized scheme. Another interesting observation is that the proposed hybrid scheme consistently delivers the best performance, irrespective of the number of antennas, and significantly outperforms both centralized and distributed schemes. This highlights its versatility for practical fronthaul-limited CF-mMIMO systems.
Figure~\ref{fig2} illustrates the performance of different precoding schemes as a function of the number of antennas per AP, $L$, for various AP counts, $M$. Distributed precoding shows significant gains with increasing $L$, while centralized and hybrid schemes exhibit a more complex behaviour due to two conflicting effects: a larger $L$ enhances macro-diversity but also increases fronthaul load for precoding vectors, thereby reducing the number of centrally served users. For $M = 20$, centralized precoding outperforms the distributed countepart when $L \leq 18$, but is overtaken beyond that point. The proposed hybrid scheme consistently delivers the best performance across all scenarios, demonstrating its robustness under fronthaul constraints in CF-mMIMO systems. This figure also highlights the adaptability of the proposed  OPA and K-means-based user grouping across diverse configurations.

%=========
%\comm{\subsection{Computational Complexity and Fronthaul Requirement }
%In hybrid precoding design, the centralized ZF precoding for users in $\kcset$ has a complexity of $\mathcal{O}(MLK_c^2+K_c^3)$, while  local ZF for $\kdset$ users is computed independently at each AP with lower per-AP complexity  $\mathcal{O}(LK_d^2+K_d^3)$.   The  centralized ZF and distributed baselines have the complexity of  $\mathcal{O}(MLK^2+K^3)$ and  $\mathcal{O}(LK ^2+K^3)$, receptively. In addition, the fronthaul usage for sending precoding vecors in hybrid scheme is per AP is $\mathrm{FH}_{m,\mathrm{pr}}=  K_\zf \alpha_2,$, while for centralizes}

 Finally, Fig.~\ref{fig3} shows the performance of the proposed hybrid  design for various $\mathrm{FH_{max}}$ when the total number of served users must be equal to $K$. Here,  $\Kmax^\zf$ is calculated as $\Kmax^\zf\!=\! (\mathrm{FH}_{\mathrm{max}}\!-\!K\alpha_2)/\alpha_1$ according to~\eqref{eq:Kmax}. We also include a channel gain-based heuristic   for user grouping, denoted as LSF,  where users are ranked based on their total channel gain, $\sum_{m=1}^M\beta_{mk}$,  and the top $\Kmax^{\zf}$
 users are selected for centralized ZF.   The results show that the K-means and LSF schemes achieve comparable SE, with K-means yielding a modest gain $3\%$ under our evaluated scenario. This suggests that both strategies are viable, and the K-means method offers a flexible and effective solution, especially in scenarios where channel strength-based ranking is less reliable or harder to implement, i.e., under non-uniform deployments, mobility conditions, or more complex channel models.

%----------------------------------------------------
\vspace{-0.7em}
\section{Conclusions}
We have proposed a hybrid centralized-distributed  precoding strategy for fronthaul-limited CF-mMIMO systems. By dividing users into two groups and utilizing both ZF-based centralized and distributed precoding schemes, our approach  flexibly adapts  to varying fronthaul and SE requirements. We formulated and efficiently solved the optimization problem for user grouping and power control.   Our numerical results demonstrated that the proposed optimization approach outperformed the benchmarks. These results confirm that the hybrid strategy is well-suited for a wide range of CF-mMIMO applications with diverse fronthaul requirements.
\vspace{-0.4em}
%=================================================
%=========================================
\appendices

%===================================================
%==========================================================================================================

%=======================================================================================================
\vspace{-0.2em}
 \bibliographystyle{IEEEtran}
 \bibliography{IEEEabrv,Ref_CellFreeFronthaul}
%==============================================================================
%==============================================================================
\end{document}